# Spin/valley coupled dynamics of electrons and holes at the MoS$_2$-MoSe$_2$ interface

Abhijeet Kumar[1], Denis Yagodkin[1], Nele Stetzuhn[1], Sviatoslav Kovalchuk[1], Alexey Melnikov[2], Peter Elliott[3], Sangeeta Sharma[3], Cornelius Gahl[1], and Kirill I. Bolotin[1,*]

[1]Department of Physics, Freie Universität Berlin, 14195 Berlin, Germany
[2]Institute for Physics, Martin Luther University Halle, 06120 Halle, Germany
[3]Max-Born-Institut für Nichtlineare Optik und Kurzzeitspektroskopie, Max-Born Straße 2a, 12489 Berlin, Germany

*kirill.bolotin@fu-berlin.de

# Abstract

The coupled spin and valley degrees of freedom in transition metal dichalcogenides (TMDs) are considered a promising platform for information processing. Here, we use a TMD heterostructure $MoS_2$-$MoSe_2$ to study optical pumping of spin/valley polarized carriers across the interface and to elucidate the mechanisms governing their subsequent relaxation. By applying time-resolved Kerr and reflectivity spectroscopies, we find that the photoexcited carriers conserve their spin for both tunneling directions across the interface. Following this, we measure dramatically different spin/valley depolarization rates for electrons and holes, ~30 $ns^{-1}$ and <1 $ns^{-1}$, respectively and show that this difference relates to the disparity in the spin-orbit splitting in conduction and valence bands of TMDs. Our work provides insights into the spin/valley dynamics of free carriers unaffected by complex excitonic processes and establishes TMD heterostructures as generators of spin currents in spin/valleytronic devices.

# Introduction

Two-dimensional materials from the group of monolayer transition metal dichalcogenides (TMDs) have emerged as promising candidates for applications in spintronics and valleytronics. In monolayers of these materials, electron-hole pairs can be selectively excited in either of the two inequivalent but energetically degenerate K and K' valleys in momentum space using circularly polarized light. At the same time, strong spin-orbit coupling in TMDs ensures the coupling of the spin and valley degrees of freedom[1,2]. Rich spintronic properties of TMDs have been confirmed in recent experiments exploring light-induced generation[3,4], spatial transport[5–7], and coherent manipulation of spin/valley polarized excitations[8–10]. In quasi-2D monolayer TMDs, the weak screening of the electron-hole interaction causes the formation of tightly bound excitons[11]. The strong electron-hole interaction also leads to quick loss of their valley polarization through decoherence caused by the exchange interaction[12], and to quick dissociation of excitons. In contrast, spin/valley polarization of free carriers in TMDs is unaffected by electron-hole recombination and exchange interaction. The spin/valley degree of freedom of these free carriers is hence expected to be a robust information carrier compared to the short-lived excitons. While it has been shown that the spin/valley polarization in monolayer TMDs is at least partially transferred from excitons to the resident carriers that are present after photoexcited carriers relax[13–17], the mechanism of this transfer is not fully understood.

Van der Waals heterostructures composed of two dissimilar TMDs represent the next logical step in complexity towards spintronic applications. For a heterostructure with type-II band alignment, such as $MoS_2$-$MoSe_2$, it is energetically favorable for an electron and a hole to reside in different materials[18]. Optical excitation of a single material in the heterostructure leads to ultrafast charge transfer across the 2D interface (e.g. electrons from $MoSe_2$ into $MoS_2$ and holes from $MoS_2$ into $MoSe_2$) within 50 femtoseconds[19]. Such spatial separation of electrons and holes suppresses both ultrafast electron-hole recombination and electron-hole exchange scattering – the problems plaguing spintronic applications of monolayer TMDs. The potential of TMD heterostructures has been demonstrated in recent experiments that showed valley-polarized interlayer excitons with lifetimes in the order of nanoseconds to microseconds via helicity-resolved photoluminescence measurements[20–23].

At the same time, probing spin/valley properties of a TMD heterostructure by detecting optical signatures of interlayer excitons is challenging. Direct optical detection of interlayer excitons is limited by their low oscillator strength that disappears almost completely for angle-misaligned heterostructures[24]. Instead, bichromatic time-resolved optical spectroscopies employing the intralayer exciton transitions as indirect probe of interlayer excitons have been used to get insight into their dynamics[25,26]. Such measurements reach sub-picosecond time resolution and, unlike photoluminescence spectroscopy, are sensitive to both population and spin dynamics of electrons and holes. However, several key questions regarding spin dynamics in heterostructures remain unanswered. First, spin lifetimes for electrons and holes have not been measured and compared in a single TMD heterostructure. This comparison is especially interesting because of the order of magnitude difference in the spin-orbit interaction strength for conduction and valence bands in TMDs. While these lifetimes have been reported for individual monolayer TMDs[14,15], the dynamics in monolayers are likely dominated by poorly understood localized or dark excitons as well as by resident carriers[13–16,27,28]. In contrast, ultrafast separation of photoexcited electrons and holes in a type-II heterostructure allows isolating the contribution of processes dominated by intralayer excitons. Similarly, despite several recent measurements showing spin conservation upon tunneling[21,29], sub-picosecond dynamics of spin/valley polarization transfer of electrons and holes across a TMD heterostructure interface have not been explored. Finally, to the best of our knowledge it is not understood if the process of spin polarization transfer onto resident carriers affects TMD heterostructures.

In order to address these open questions, we apply ultrafast two-color transient reflection and Kerr rotation microscopies with sub-200 fs time resolution. By resonantly exciting one layer of the $MoS_2$-$MoSe_2$ heterostructure with circularly polarized light and probing the transient signal in the other layer, we access the ultrafast spin/valley polarization and population dynamics of electrons and holes. We find that the electrons and holes preserve their spin upon transfer across the heterostructure interface. The spin/valley signal in the latter layer has an apparent build-up time of 150−300 fs, significantly slower than the charge transfer time (<40 fs) for both electrons and holes. In contrast, we observe distinct relaxation dynamics for electrons and holes in the heterostructure: while electron spins in $MoS_2$ depolarize fast, over tens of picoseconds, the spin/valley lifetime of holes in $MoSe_2$ is limited by their population decay. We attribute this to different spin scattering rates for electrons and holes, which, in turn, originate from the different magnitude of spin-orbit splitting between the valence and conduction bands in TMDs. We also find that the process of spin polarization transfer to resident carriers is not effective in our heterostructure.

## Results

### $MoS_2$-$MoSe_2$ heterostructure

From the family of TMD materials, we chose a $MoS_2$-$MoSe_2$ heterostructure. The two components are picked to be similar (Mo-based), with comparable spin-orbit interaction strengths[30] and exciton binding energy[31]. The schematic spin texture of a Mo-based monolayer TMD near K/K' points in the Brillouin zone is shown in Fig. 1a. Spin-orbit interaction leads to strong spin-splitting of the valence bands (147 meV for $MoS_2$ and 186 meV for $MoSe_2$) and weak splitting of the conduction bands (3 meV for $MoS_2$ and 21 meV for $MoSe_2$) both at K and K' valleys[30]. Circularly polarized excitation near resonance with the optical bandgap of the TMD ("A transition") couples to the transition between the lower conduction sub-band and the higher valence sub-band either in K or K' valley, depending on the excitation chirality. Therefore, both the spin and valley degree of the photoexcited electron/hole pair are defined by the excitation chirality. As the $MoS_2$-$MoSe_2$ heterostructure exhibits a type II alignment[32], upon resonant excitation of MoSe$_2$ (MoS$_2$), photoexcited holes (electrons) stay inside the original layer, while electrons (holes) can tunnel into lower-energy states of the other material (Fig. 1b). Our main goals are to explore the conservation of spin/valley degrees of freedom upon tunneling and to study their depolarization dynamics afterwards.

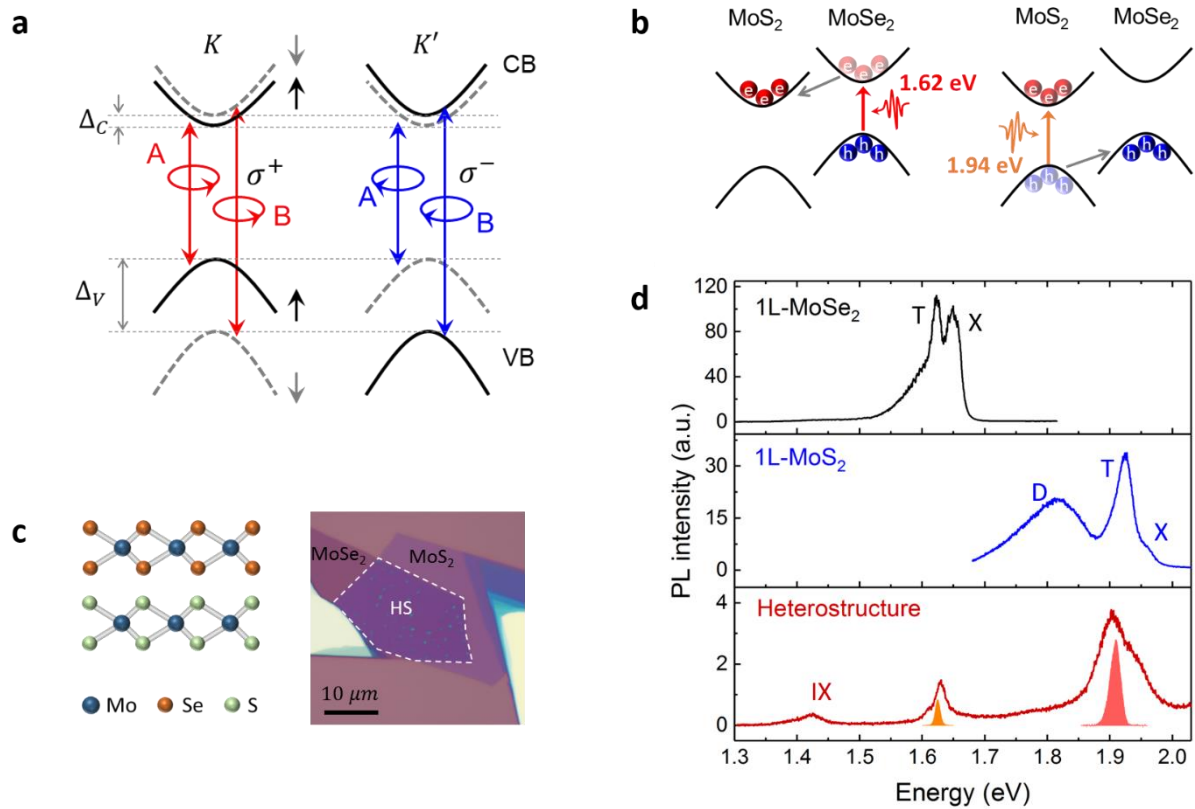

*Figure 1: TMD spin texture and heterostructure characterization. (**a**) Spin texture in a Mo-based monolayer TMD. Coupled spin/valley degree of freedom can be selectively excited using circularly polarized light. (**b**) The response of a MoS$_2$-MoSe$_2$ heterostructure to photoexcitation. Following light absorption by individual layers, electrons and holes tunnel into energetically favourable states in MoS$_2$ and MoSe$_2$, respectively. (**c**) Left: cartoon view of the heterostructure. Right: optical image of the assembled device on a SiO$_2$/Si substrate. Dashed white line shows the heterostructure boundary. (**d**) Photoluminescence spectra of monolayer MoSe$_2$ (top panel), monolayer MoS$_2$ (middle panel), and the heterostructure (bottom panel) at 5 K. Significantly quenched emission from the heterostructure along with an interlayer exciton peak (IX) at 1.42 eV suggest efficient charge transfer across the interface. Two shaded peaks in the bottom panel, red at 1.62 eV and orange at 1.91 eV, show the spectra of pump/probe beams tuned to MoSe$_2$ and MoS$_2$ optical bandgap, respectively.*

Figure 1c is an optical image of one MoS$_2$-MoSe$_2$ device (fabrication details are in Methods). In addition to the heterostructure region (marked by dashed white line), the device also contains areas of monolayer MoS$_2$ and monolayer MoSe$_2$. Low temperature (5 K) photoluminescence (PL) spectra (Fig. 1d) from these monolayer regions exhibit neutral (X) and charged (T) exciton peaks, at 1.65 eV and 1.62 eV for MoSe$_2$ (top panel) and at 1.96 eV and 1.93 eV for MoS$_2$ (middle panel), respectively[33,34]. In addition, the monolayer MoS$_2$ also exhibits broad defect-assisted emission (D) at 1.8 eV[35]. The PL spectrum from the heterostructure region (bottom panel) contains the peaks corresponding to both MoSe$_2$ and MoS$_2$, as expected. These peaks are redshifted compared to those of the isolated monolayers, likely due to modified screening[36].

At the same time, the PL intensity of both materials is significantly quenched suggesting efficient charge transfer across the 2D interface[19,37]. In addition, a new peak at 1.42 eV appears in the heterostructure region. This peak, a tell-tale sign of type-II heterostructures, originates from interlayer exciton (IX) emission and indicates strong interlayer coupling. We confirm that the spectral features of both monolayers in the heterostructure are well separated and can be pumped and probed selectively.

**Ultrafast spin/valley polarization transfer of electrons and holes**

We now turn to the ultrafast Kerr and transient reflectivity measurements. The setup for pump-probe Kerr rotation and transient reflectivity measurements is depicted in Fig. 2a (details are in Methods). The photon energies of pump and probe, shown as red and orange shaded regions in Fig. 1d (bottom panel), were tuned to the excitonic resonances of the respective materials. The transient reflectivity signal for probe in resonance with the A transition corresponds to the total pump-induced population change in the relevant sub-bands at K and K' valleys[38] (Fig. 1a). At the same time, the Kerr rotation angle, $\theta_K$, measures the polarization axis rotation of the reflected probe, and is roughly proportional to the population imbalance between the same sub-bands at K and K' valleys[17]. The population imbalance between K and K' valleys of higher conduction sub-bands or lower valence sub-bands can also be accessed by tuning the Kerr probe wavelength to the resonance with the B transition (Fig. 1a).

We first focus on the spin/valley and population dynamics in the MoS$_2$-MoSe$_2$ heterostructure during the first picosecond following photoexcitation in one of the layers. Fig. 2b shows the time-resolved Kerr signal under excitation at the MoSe$_2$ optical bandgap (hv$_{pump}$ = 1.62 eV) and probe at the MoS$_2$ optical bandgap (hv$_{probe}$ = 1.91 eV) for different pump polarization. In this case, excited electrons can tunnel from MoSe$_2$ to MoS$_2$. The transient Kerr signal is odd with respect to the pump chirality, $\theta_K(\sigma^+) = -\theta_K(\sigma^-)$, and vanishes for linear pump polarization. To confirm that the transient Kerr signal indeed originates from pump-induced charge transfer across the heterostructure interface, we examine a spatial map of the Kerr signal intensity, $\theta_K(\sigma^+) - \theta_K(\sigma^-)$, at 2 ps delay (Fig. 2c). The signal is uniform across the heterostructure, while no Kerr signal is observed from monolayer regions. Switching the photon energies of pump to 1.94 eV and probe to 1.62 eV, we access the case of holes tunneling from MoS$_2$ to MoSe$_2$. Also in that case, the Kerr signal is largely uniform across the heterostructure, and vanishes in monolayer areas (Supplementary Fig. 2d, 2e). In Fig. 2d, we compare the time evolution of the reflectivity (gray traces) and Kerr (blue traces) signals in MoSe$_2$ (sensitive to tunneled holes, upper panel in Fig. 2d) and in MoS$_2$ (sensitive to electrons, lower panel in Fig. 2d), respectively, within the first 1.5 ps after optical excitation. For both excitation schemes, the transient reflectivity signal rises on the timescale of our time resolution. The rise of Kerr signal, in contrast, is delayed and reaches its maximum at a delay of 1 to 2 ps.

The appearance of a transient reflectivity signal in one material of the heterostructure after the excitation of the other material could result from the transfer of energy[39] or charge[18] across the MoS$_2$-MoSe$_2$ interface. Energy transfer is forbidden in the case of pumping the lower bandgap material MoSe$_2$. For the case of pumping MoS$_2$, energy transfer in principle is allowed. However, since energy transfer is unlikely to conserve spin, the main effect of this process on the Kerr signal would be a reduction of our total Kerr amplitude without changing the picosecond dynamics. Additionally, any possible energy transfer would be immediately followed by fast charge transfer bringing electrons back into MoS$_2$ and holes into MoSe$_2$. We conclude that the transient optical signals are dominated by photoexcited holes transferred across the heterostructure interface upon pumping $MoS_2$, and electrons upon pumping $MoSe_2$.

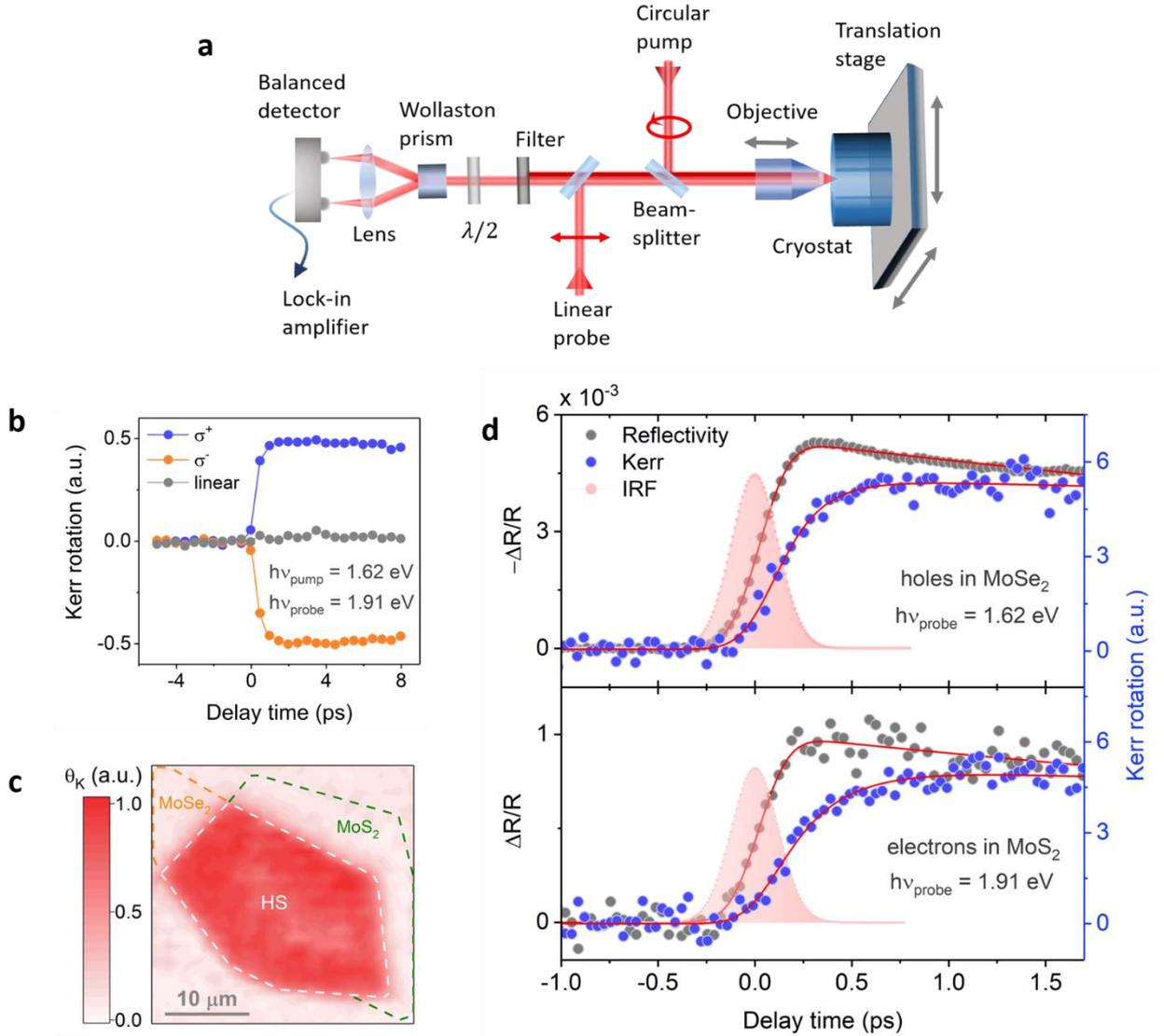

*Figure 2: Charge and spin/valley dynamics in $MoS_2$-$MoSe_2$ on the picosecond timescale. (a) Time-resolved Kerr/reflectivity setup. (b) Kerr rotation dynamics for the case of pumping $MoSe_2$ and probing $MoS_2$. The odd nature of Kerr signal with pump chirality indicates spin/valley polarization in $MoS_2$. (c) Spatial map of Kerr signal at 2 ps time delay under the same experimental conditions as b). The appearance of Kerr signal only from the heterostructure region confirms that it originates from the transfer of spin-polarized electrons across the heterostructure interface. (d) Evolution of transient reflectivity (gray) and Kerr signal (blue) for holes (upper panel) and electrons (lower panel) together with the cross-correlation of the two laser pulses (shaded red). Fits to the data are shown as solid red lines.*

The appearance of Kerr signal for probe resonant with the optical bandgap (Fig. 2b-d) indicates an imbalance between K and K' valleys of the lower conduction sub-bands for electrons and higher valence sub-bands for holes (Fig. 1a). However, due to the band offset between $MoS_2$ and $MoSe_2$[32], photoexcited carriers possess excess kinetic energy and can relax into higher energy states, that is, the higher conduction sub-bands for electrons and lower valence sub-bands for holes. These sub-bands have opposite spin texture and do not contribute to the Kerr signal for probe resonant with the optical bandgap. To accurately account for spin contribution from higher energy sub-bands, we note that the Kerr signal for the probe energy resonant with the B transition in $MoS_2$ is sensitive to the K/K' population imbalance of higher-lying conduction sub-bands of $MoS_2$ (Fig. 1a). We observe that the Kerr signal probed at the B transition flips its sign with respect to the previously considered case of probing at the optical bandgap (Supplementary Fig. 1a, 1b). This indicates that the valley polarization of higher sub-bands is opposite compared to that of lower sub-bands. Therefore, we conclude that the spin degree of freedom of the tunneled carriers is conserved[21,29], consistent with previous reports of charge transfer being faster than spin/valley depolarization of intralayer excitons[40].

To quantify the dynamics of charge and spin transfer across the heterostructure, we fit the data in Fig. 2d with the sum of an exponential rise and a slow exponential decay function convolved with the instrument response function (IRF). From the fit, we estimate the charge transfer time to be less than 40 fs for both electrons and holes, consistent with previous reports[19,25,41]. For the delayed evolution of the Kerr signal, we extract time constants of $250 \pm 70$ fs for electrons and $170 \pm 40$ fs for holes (Supplementary note 1). The origin of this discrepancy between the reflectivity and Kerr signal rise times is not clear. We speculate that the difference in rise times is related to the cooling dynamics of hot tunneled carriers. Indeed, while the reflectivity signal originates at least in part from phenomena such as screening that are insensitive to the energy of the carriers[42], the Kerr signal is caused by carrier population imbalance close to the extrema of conduction and valence bands at the K/K' points[4]. Therefore, the delayed rise of the Kerr signal may reflect the relaxation dynamics of hot photoexcited carriers towards the band edges after tunneling. It has been suggested that hot carrier relaxation does not cause valley depolarization[8]. Our observation is consistent with a recent study suggesting that hot carriers can relax to the band edges on sub-picosecond time scale via emission of phonons[43].

**Dynamics of electrons and holes after 1ps**

The decay of photoexcited electron and hole population as well as depolarization of their spin/valley degree of freedom proceeds on timescales longer than 1 ps. Figure 3 is the central result of our paper comparing the reflectivity (gray trace) and Kerr signals (blue trace) for the cases of pumping $MoS_2$ and probing $MoSe_2$ (Fig. 3a) and vice versa (Fig. 3b). As explained before, the reflectivity signal can be interpreted as the time-dependent density of photoexcited carriers, the Kerr signal as spin/valley population imbalance; the case of probing $MoSe_2$ corresponds to detecting tunneled photoexcited holes in $MoSe_2$, while the reverse case corresponds to electrons in $MoS_2$.

The population decay dynamics are similar between holes (Fig. 3a, gray points) and electrons (Fig. 3b, gray points). The hole population dynamics are fitted with a bi-exponential decay with fast $\tau_1^{h,p} = 30 \pm 2$ ps and slow $\tau_2^{h,p} = 420 \pm 40$ ps components. For electrons, we observe similar dynamics with components $\tau_1^{e,p} = 63 \pm 7$ ps and $\tau_2^{e,p} = 676 \pm 75$ ps. In contrast to the population dynamics, the spin/valley dynamics differ greatly between electrons and holes. For holes, we again observe a bi-exponential decay with fast $\tau_1^{h,s} = 38 \pm 3$ ps and slow $\tau_2^{h,s} = 640 \pm 70$ ps components (Fig. 3a, blue trace). For electrons, in contrast, we observe a fast mono-exponential decay with a much shorter lifetime $\tau^{e,s} = 28 \pm 1$ ps (Fig. 3b, blue trace).

We speculate that similar lifetimes of electron and hole population in our heterostructure reflect electron/hole recombination across the interface. It is interesting to note that while we do observe the emission from IX excitons in our heterostructure (Fig. 1d), much longer lifetimes in the nanosecond range have been measured for these excitons via time-resolved PL measurements compared to hundreds of picosecond population lifetimes seen here[20,21,23]. This suggests that the recombination may be dominated by a defect-assisted nonradiative pathway, also predicted in a recent report[25]. The observation of somewhat different lifetimes for electrons and holes as well as of slow and fast decay components may reflect the difference in the defect densities between the two materials and spatially varying disorder at the heterostructure interface[20].

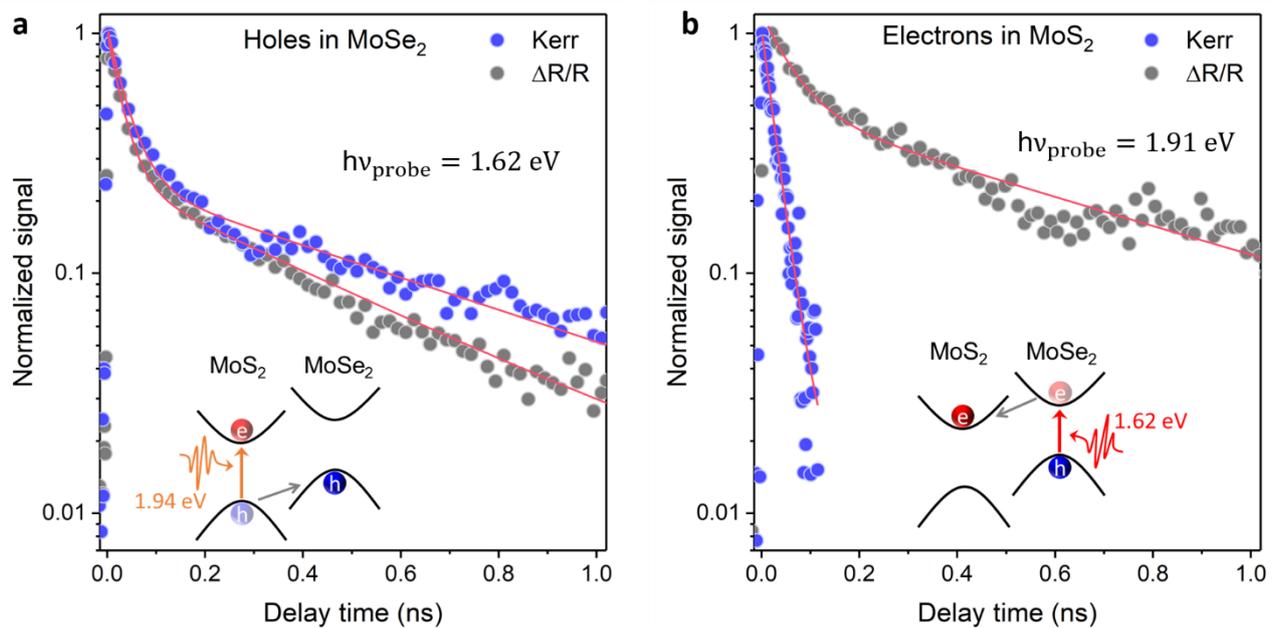

*Figure 3: Charge and spin/valley dynamics of electrons and holes in MoS$_2$-MoSe$_2$ on 1 ns timescale. (a) Time-resolved dynamics for the case of pumping MoS$_2$, probing MoSe$_2$. These traces are interpreted as the dynamics of carrier population (gray) and spin/valley population imbalance (blue) for holes in MoSe$_2$. (b) Corresponding case of pumping MoSe$_2$ and probing electrons in MoS$_2$. Red lines are fits to the data.*

It is important to analyze confounding effects other than spin/valley- and charge- relaxation that could potentially contribute to the data in Fig. 3. First, as noted earlier, a pump resonant with the MoS$_2$ bandgap also non-resonantly excites the lower-bandgap material MoSe$_2$. However, our signals are not significantly affected by intralayer dynamics in MoSe$_2$. This is confirmed by i) the lack of Kerr signal for non-resonant excitation of monolayer MoSe$_2$ (Supplementary Fig. 2d, 2e) and ii) approximately 7 times weaker reflectivity signal in monolayer MoSe$_2$, that vanishes within 20 ps after photoexcitation compared to the heterostructure region (Supplementary Fig. 2b, 2c). In addition, we do not observe any dependence of the reflectivity dynamics on pump fluence (Supplementary Fig. 3) which would be expected for intralayer dynamics in TMDs[44,45]. Second, laser-induced changes in temperature of the heterostructure could also change the relaxation dynamics. Such temperature-related effects are also ruled out by the lack of pump fluence dependence. Finally, photoexcited carriers could potentially diffuse out of the probe spot during their lifetime. However, this effect is likely insignificant in our system since we can estimate the diffusion length to be <10 nm within our observed lifetime[46].

**Discussion**

The dramatically different nature of spin/valley dynamics of holes in MoSe$_2$ and electrons in MoS$_2$ (blue symbols in Fig. 3a and 3b) is the most striking feature of our data. While a significant part of hole spin polarization lives as long as the carrier population (several hundred picoseconds) the lifetime of electron spins is more than one order of magnitude shorter compared to both electron and hole population lifetimes.

In general, the measured spin/valley lifetime may originate from two different mechanisms. First, it may be limited by the processes leading to the loss of spin/valley polarization, such as intervalley and spin-flip intravalley scattering. Second, it may simply result from the decay in the population of excited carriers even in the absence of spin polarization loss. To distinguish between these two mechanisms limiting spin/valley lifetime, we introduce the degree of spin/valley polarization

$$P = \frac{P_+ - P_-}{P_+ + P_-}$$

Here $P_+$ and $P_-$ are the transient population densities in the lowest-energy sub-bands of K and K' valleys, respectively. Experimentally, we can determine $P$ by noting that our transient reflectivity signal is approximately proportional to the total population of excited carriers ($P_+ + P_-$), while the Kerr signal is proportional to the valley population imbalance ($P_+ - P_-$). Figure 4 shows $P$ determined from the ratio of Kerr and reflectivity signals. For electrons (Fig. 4, red trace), we observe a mono-exponential decay of $P$ with a decay time of 30 ps, identical to what is seen in Fig. 3b. In contrast, the degree of spin/valley polarization for holes remains almost constant, within uncertainty, over our delay time window of 1 nanosecond (Fig. 4, blue trace). This means that the decay of the Kerr signal for holes in MoSe$_2$ seen in Fig. 3a is fully caused by the loss of carrier population and the intrinsic rate of hole valley depolarization is significantly lower than 1 ns$^{-1}$.

Summarizing our interpretations so far, the fast Kerr signal decay (<30 ps) observed in MoS$_2$ is determined by the fast spin/valley depolarization of electrons in the conduction band, while the much slower Kerr signal decay (>600 ps) measured in MoSe$_2$ is determined by the population lifetime of photoexcited holes. Similar differences between electron and hole spin lifetimes were found across all measured samples (Supplementary Table 1), which confirms, consistent with previous reports[25,29] that stacking angle does not significantly influence spin transfer and relaxation dynamics across the interface. The difference between spin/valley lifetimes for electrons and holes is noteworthy since most of the material parameters of MoS$_2$ and MoSe$_2$ including electron and hole effective masses are similar. However, the magnitude of spin-orbit splitting is very different between the conduction and the valence bands of both materials. A small splitting of ~3 meV in the MoS$_2$ conduction band facilitates rapid intravalley and intervalley scattering of electrons[30]. Spin-flip intravalley scattering may arise due to magnetic impurities or phonon-assisted processes (e.g. via phonons with near zero momentum[47]). Intervalley scattering of electrons may be mediated by longitudinal acoustic (LA) phonons[48–50] leading to rapid spin/valley depolarization. In contrast to electrons, both intra- and intervalley scattering of holes are suppressed due to the spin-splitting of the MoSe$_2$ valence band, ~186 meV, which is much larger than available phonon energies[30]. We therefore suggest that the difference between spin/valley depolarization rates for electrons and holes is related to the difference in the strength of spin-orbit interactions in conduction and valence bands of TMDs.

The arguments above also show that the process of intervalley scattering may effectively render the electrons in $MoS_2$ "invisible" for probing at the optical bandgap by scattering them into the higher conduction sub-bands. To analyze the spin contribution from that sub-band, we probe the decay of Kerr signal at the energy corresponding to the B transition in $MoS_2$. As explained earlier, this signal reflects the K/K' valley population imbalance between electrons in the higher-lying conduction sub-bands of $MoS_2$. The measured decay dynamics of this signal are identical to the dynamics probed resonantly with the optical bandgap (Supplementary Fig. 1c). This confirms that the spin-polarization of electrons decays at the same rate in both sub-bands that is likely determined by the efficiency of spin-flip intravalley scattering. Overall, we conclude that the measured decay dynamics of *P* (Fig. 4) reflects the decay of spin polarization of all excited electrons and all excited holes, with rates 30 ns$^{-1}$ and <1 ns$^{-1}$, respectively. To the best of our knowledge, this is the first direct comparison of spin lifetime of electrons and holes in a TMD heterostructure.

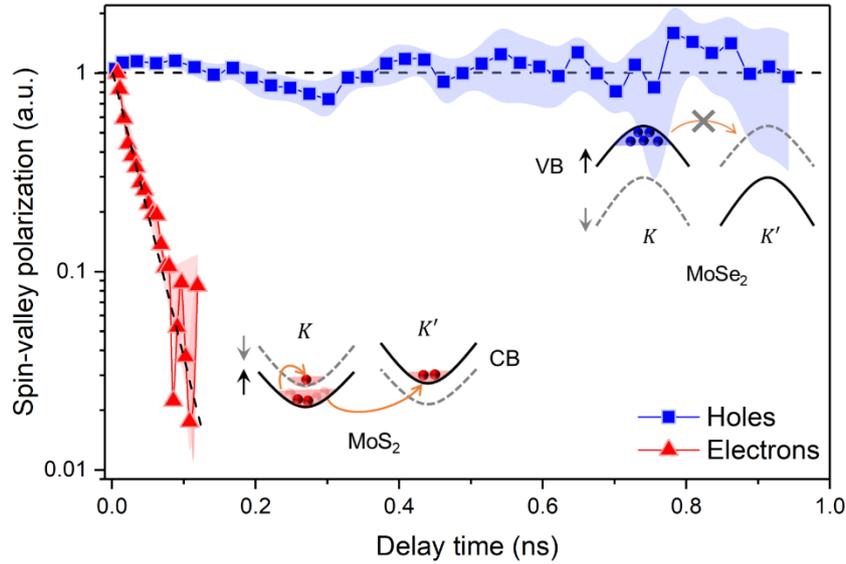

*Figure 4: Spin/valley polarization for electrons and holes. Spin/valley polarization dynamics for electrons in $MoS_2$ (red triangles) and holes in $MoSe_2$ (blue squares). Electrons in the lower conduction sub-band in K/K' valley can undergo rapid intravalley and intervalley scattering, contributing to the equilibration of the valley polarization of the lower sub-bands (inset, bottom). Holes retain their spin/valley polarization over a delay range of 1 ns owing to suppressed intervalley scattering (inset, right). The signal is normalized with respect to the spin/valley polarization at zero delay. Dashed black line is a guide to the eye.*

Strikingly different electron and hole spin lifetimes in our heterostructure are consistent with recent measurements of spin lifetimes in a doped monolayer TMD[14,15]. The nature of the state exhibiting the long spin lifetime in these monolayers is still debated, however. It has been suggested that the spin polarization of relatively short-lived intralayer excitons is transferred onto other long-lived states including resident carriers[13,14,17,42,51], localized states[15], or dark excitons[27,28]. In contrast, in a type-II heterostructure employed here, electrons and holes are separated between the layers and various states complicating the interpretation of monolayer measurements do not arise. For example, while the decay dynamics of the excited electrons in the higher and lower conduction sub-bands of MoS$_2$ are identical in our measurements, the presence of dark excitons in W-based TMDs leads to dramatically different dynamics of these excitations within the same sub-bands[42]. In addition, the lack of Kerr signal after the decay of the excited population (unlike in the measurements of Ref. 42, 51) suggests that the process of spin transfer onto resident carriers is not effective in our devices. We also note that very long valley lifetimes in the microsecond range were reported for holes in similar heterostructures[38,52] (these works did not analyze the difference between electron and hole spin/valley dynamics). We speculate that significantly longer lifetimes compared to our measurements may be related to the different measurement scheme of Ref. 38, 52, where the same TMD layer was excited and probed. As a result, photoexcited holes never tunneled into the other layer, unlike in our experiment. Nevertheless, the tunneled holes in our heterostructure maintain their spin/valley polarization within their population lifetime.

## Conclusion

To summarize, we investigated the coupled spin/valley dynamics of electrons and holes in MoS$_2$-MoSe$_2$ heterostructures. Using a combination of transient Kerr and reflectivity microscopies, we showed that the photoexcited electrons and holes cross the heterostructure interface within 40 fs while largely preserving their spin. We then studied the decay of spin polarization of the photoexcited carriers and found very different dynamics for holes in MoSe$_2$ and electrons in MoS$_2$. For electrons, the rapid decay of spin/valley polarization with a time constant of 30 ps is related to fast intravalley and intervalley scattering in the conduction band of MoS$_2$. For holes, in contrast, the observed decay of Kerr signal is related to the loss of their population, while the rate of spin/valley depolarization is lower than 1 ns$^{-1}$. The difference in the rates of spin/valley depolarization between electrons and holes is explained by the disparity of spin-orbit splitting between conduction and valence bands in TMDs. The process of spin transfer to resident carriers is not effective in our heterostructures. To the best of our knowledge, this is the first study reporting the spin/valley lifetimes of electrons and holes in a TMD heterostructure. Our results are relevant for the emerging use of TMD heterostructures as sources of spin/valley polarized currents in layered materials. Finally, we note that it will be interesting to investigate the initial stages of the spin/charge transfer across the TMD interface with techniques capable of resolving sub-100 fs dynamics. It will be equally interesting to analyze the effect of defects, which likely provide the momentum needed for the carriers to cross the TMD interface.

## Methods

**Sample preparation**
To fabricate the device, individual MoS$_2$ and MoSe$_2$ flakes were mechanically exfoliated on a PDMS (polydimethylsiloxane) substrate and subsequently transferred onto a SiO$_2$/Si substrate using a well-established dry-transfer technique[53] without controlling the stacking angle. The heterostructure samples were vacuum annealed at 200°C for 30 minutes to improve interlayer coupling. Overall, 6 samples were fabricated and measured.

**Pump-probe spectroscopy**
The samples placed inside a cryostat were measured at a base temperature of 4.2 K, unless stated otherwise. The cryostat was mounted on a motor-controlled xy-translation stage with sub-micrometer spatial resolution. We used a wavelength-tunable femtosecond pulsed laser system (Coherent Chameleon Ultra II + compact OPO-VIS) with sub-200 fs pulse duration and 80 MHz repetition rate. The photon energies of pump and probe, shown as red and orange shaded regions in Fig. 1d (bottom panel), were tuned to the excitonic resonances of the respective materials. Fine tuning in the pump/probe energy was done to optimize the Kerr signal. Both pump and probe pulses were focused into a roughly 1 $\mu m^2$ spot on the sample at normal incidence. The pump fluence was varied between 10 $\mu J/cm^2$ and 100 $\mu J/cm^2$ in our experiments. The reflected linearly polarized probe was passed through an optical bridge consisting of a half-wave plate and a Wollaston prism, which separates the beam into two components of orthogonal polarization. Both signals were recorded by using a balanced photodetector and a lock-in amplifier phase-locked to a chopper in the pump path.

**Low T PL measurements**

For low T PL measurements, we used another optical setup (Witec Alpha confocal spectroscopy setup) with a Nikon 50×SLWD objective of NA = 0.5. Samples were placed in a cryostat at 5 K. The laser excitation wavelength was set to 532 nm and 633 nm to access the spectral range of $MoS_2$ and $MoSe_2$, respectively. The laser power was varied within 20 $\mu$W and 30 $\mu$W.

## Acknowledgements
We acknowledge the German Research Foundation (DFG) for financial support through the Collaborative Research Center TRR 227 Ultrafast Spin Dynamics and ERC Starting grant no. 639739. We thank Ben Weintrub for technical help.

## Author contributions
K.I.B., and C.G. conceived the project; K.I.B., C.G., and A.M. designed the experiment; A.K., D.I., N.S., and S.K. fabricated and characterized the samples; A.K., D.I., and N.S. performed the experiments; A.K. analyzed the data; A.K., K.I.B., and C.G. wrote the paper with contributions from all other co-authors.

## Competing interests
The authors declare no competing interests.

## Materials and correspondence
Correspondence and requests for materials should be addressed to K.I.B.

**Supplementary information**

# Spin/valley coupled dynamics of electrons and holes at the MoS$_2$-MoSe$_2$ interface

Abhijeet Kumar[1], Denis Yagodkin[1], Nele Stetzuhn[1], Sviatoslav Kovalchuk[1], Alexey Melnikov[2], Peter Elliott[3], Sangeeta Sharma[3], Cornelius Gahl[1], and Kirill I. Bolotin[1,*]

[1]Department of Physics, Freie Universität Berlin, 14195 Berlin, Germany
[2]Institute for Physics, Martin Luther University Halle, 06120 Halle, Germany
[3]Max-Born-Institut für Nichtlineare Optik und Kurzzeitspektroskopie, Max-Born Straße 2a, 12489 Berlin, Germany

*kirill.bolotin@fu-berlin.de

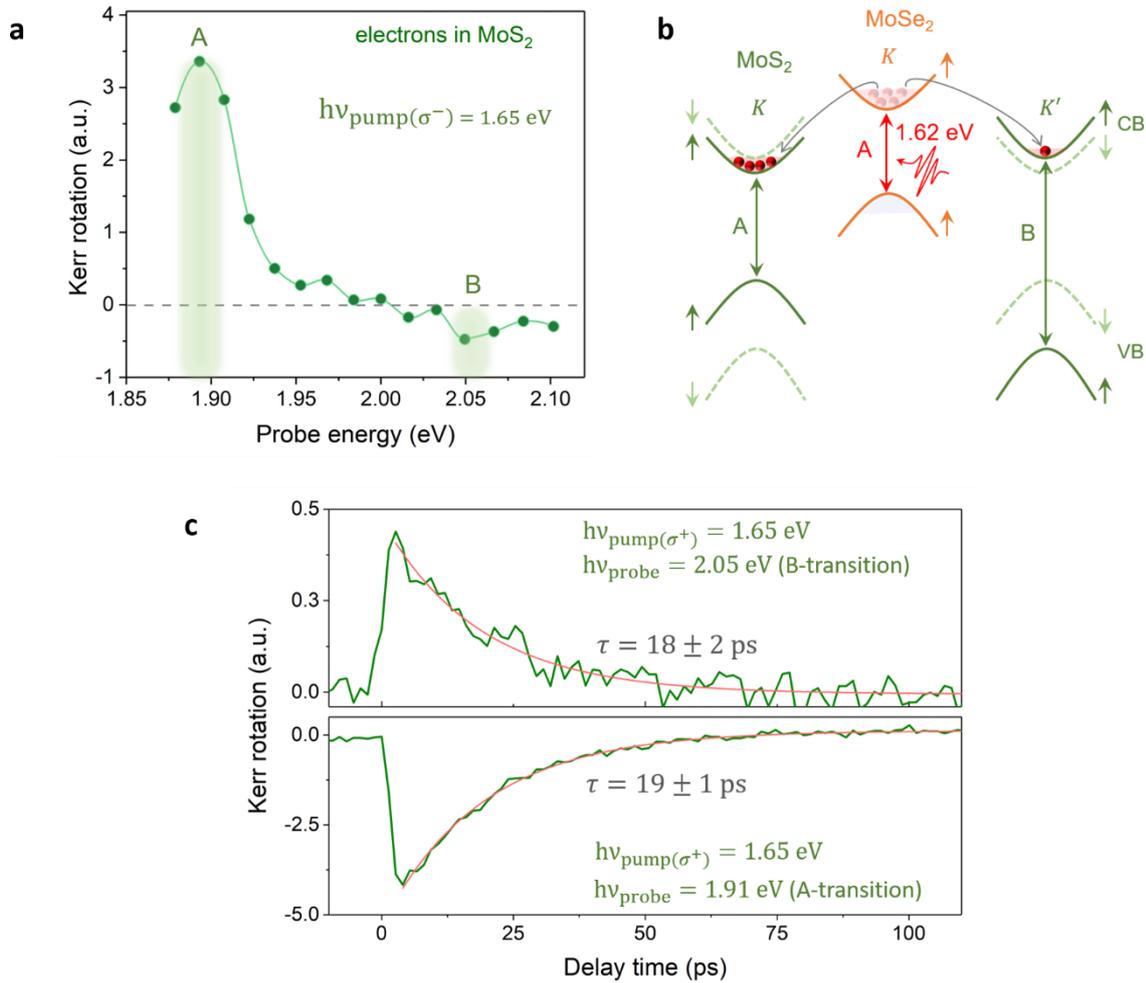

**Supplementary figure 1: Comparison of spin/valley dynamics between conduction sub-bands of MoS$_2$.** **(a)** Kerr rotation angle vs. probe photon energy (in the region of MoS$_2$ absorption) at 2 ps delay for pump corresponding to the MoSe$_2$ optical bandgap (1.65 eV). The heterostructure sample is different from the one presented in the main text. Kerr signal flips its sign between probe energies corresponding to the transition into lower conduction sub-band (A transition, ~1.90 eV) and the higher conduction band sub-band (B transition, ~2.05 eV) in MoS$_2$. **(b)** Schematic diagram for tunneling of spin-polarized electrons from MoSe$_2$ into MoS$_2$ conduction bands. **(c)** Time-resolved Kerr dynamics probed near B transition (upper panel) and near A transition (lower panel) in MoS$_2$. Similar dynamics of both traces suggest equally fast depolarization of electrons in the higher and the lower conduction sub-bands.

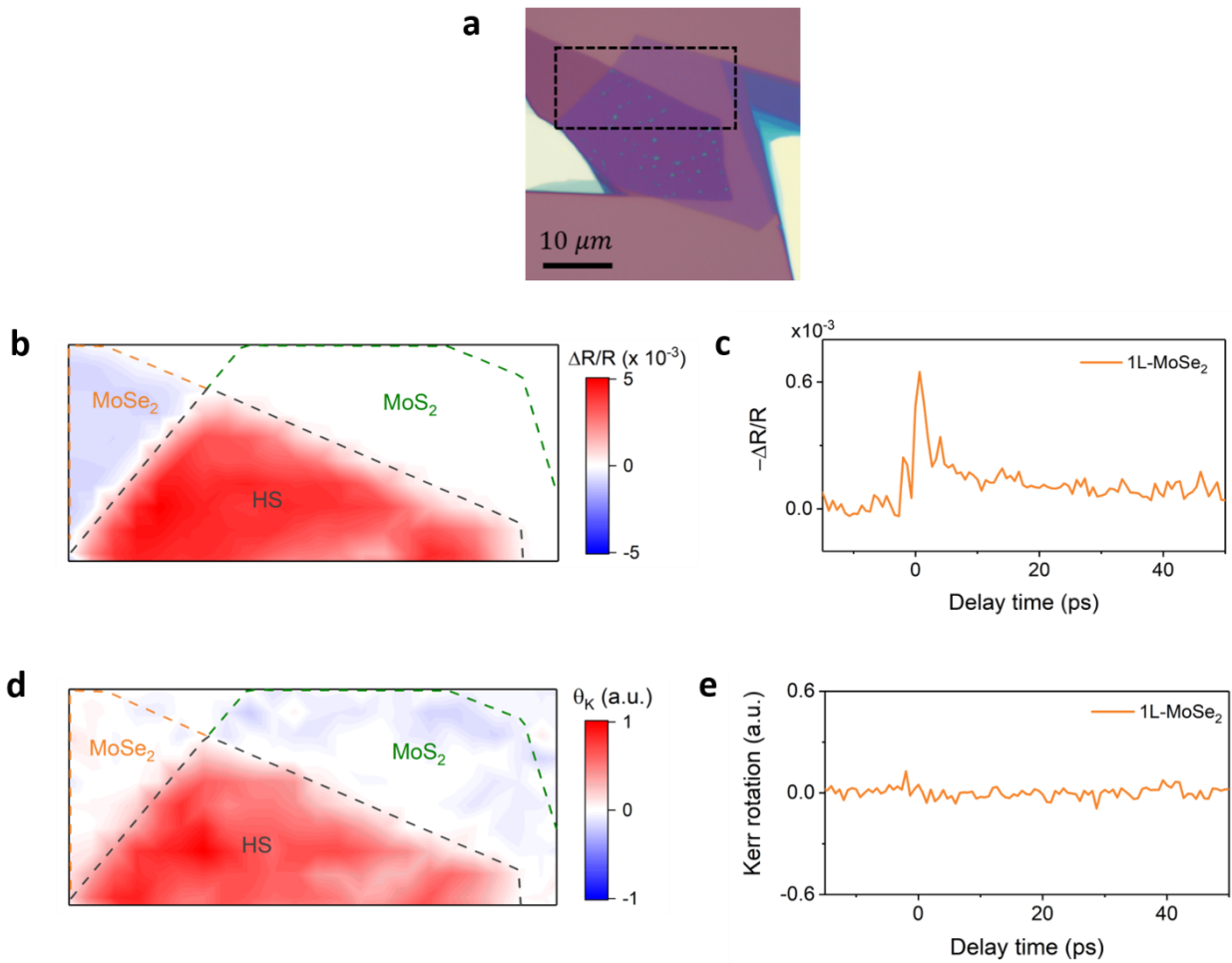

**Supplementary figure 2: Spin/valley and population dynamics in MoSe$_2$ under non-resonant excitation.** (**a**) Optical image of MoS$_2$-MoSe$_2$ heterostructure (same device as in the main text). Dashed line marks the region of maps in b) and d). Kerr and reflectivity maps/traces were recorded under pumping at 1.94 eV (optical bandgap of MoS$_2$) and probing at 1.62 eV (optical bandgap of MoSe$_2$). In addition to the excitation of MoS$_2$ and probing of MoSe$_2$ considered in the main text, these conditions also result in non-resonant excitation of lower bandgap material MoSe$_2$. (**b**) Spatial reflectivity map across the region marked in a) at 2 ps delay. Note the complete lack of the reflectivity signal from the monolayer MoS$_2$ region and approximately seven times weaker signal with opposite sign from the monolayer MoSe$_2$ region compared to the heterostructure. (**c**) Time-resolved reflectivity trace from the isolated MoSe$_2$ region. This signal is weak and decays completely within 20 ps. (**d**) Corresponding Kerr map. Note that no Kerr signal is observed from the monolayer regions. (**e**) Time-resolved Kerr trace from the isolated monolayer MoSe$_2$ region. No dynamics are observed.

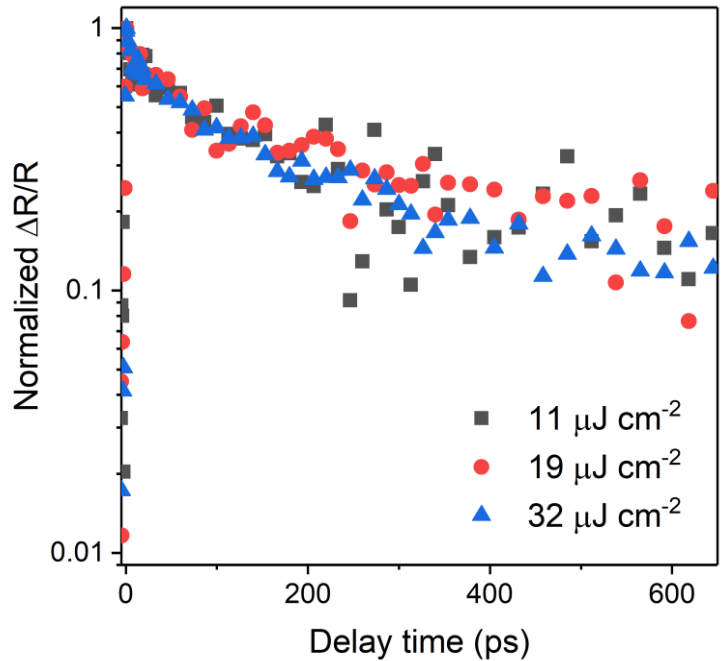

**Supplementary figure 3: Pump fluence dependence of hole population dynamics.** Hole population dynamics in the $MoS_2$-$MoSe_2$ heterostructure for three different pump fluences. Pump energy is fixed to 1.88 eV, while the probe energy is tuned to 1.59 eV, near optical bandgap of $MoSe_2$. No significant variation on the population dynamics of holes is observed. This rules outs the contribution of intralayer dynamics in $MoSe_2$, which is expected to be significantly affected by exciton-exciton annihilation mechanism.

# Supplementary note 1: Upper limit of reflectivity and spin/valley polarization rise time

We approximate the overall dynamics in our heterostructure as that of a 3-level system where a pump pulse excites population from ground state-0 to excited state-1. This is followed by ultrafast charge transfer of the excited population (electrons or holes) from excited state-1 into the energy state-2, whose decay we probe in our system. Based on this approximation, we fit the dynamics with the convolution of two functions as following:

$$f(t) = \left[-A \exp\left(-\frac{t}{\tau_{rise}}\right) + A \exp\left(-\frac{t}{\tau_{decay}}\right)\right] * \frac{1}{\sigma\sqrt{2\pi}} \exp\left(-\frac{t^2}{2\sigma^2}\right) \quad \text{(SI 1)}$$

Here, the last term in (SI 1) is the instrument response, that is, the pump-probe correlation function. The full-width at half maximum of the pump-probe correlation function is defined as $XC = 2\sigma\sqrt{2\ln(2)}$. The value of XC at the output of our laser system is 270 fs. However, we expect a slight temporal broadening of the pulses at the sample position due to significantly long beam path through several optical elements. Expecting this deviation, our aim is to estimate the upper limit of rise time for Kerr and reflectivity signal. Assuming direct population transfer from state-0 to state-2, we first fit the reflectivity signal with a single exponential decay convolved with the Gaussian cross-correlation function and extract the upper limit of XC = 286 fs. A fit to the Kerr signal using the fit function (SI 1) with XC = 286 fs gives us the upper limit of spin-valley polarization build-up time to be 170 ± 40 fs for holes and 250 ± 70 fs for electrons, respectively, which are almost identical within uncertainty. By varying the XC parameter within reasonable limits in our fitting procedure, we estimate the upper limit of reflectivity rise time to be 40 fs for both electrons and holes.

| Device # | Stacking (top-bottom) | Electron spin lifetime | Hole spin lifetime |
| --- | --- | --- | --- |
| 1 | $MoS_2$-$MoSe_2$ | 13 $\pm$ 2 ps | 39 $\pm$ 14 ps; 180 $\pm$ 8 ps |
| 2 | $MoSe_2$-$MoS_2$ (main text) | 28 $\pm$ 1 ps | 38 $\pm$ 3 ps; 640 $\pm$ 70 ps |
| 3 | $MoS_2$-$MoSe_2$ ($F_4$TCNQ doped) | 54 $\pm$ 1 ps | 13 $\pm$ 3 ps; 179 $\pm$ 14 ps |
| 4 | $MoS_2$-$MoSe_2$ | 37 $\pm$ 2 ps | 35 $\pm$ 3 ps; 830 $\pm$ 53 ps |
| 5 | $MoS_2$-$MoSe_2$ | 42 $\pm$ 3 ps | -- |
| 6 | $MoSe_2$-$MoS_2$ | 19 $\pm$ 1 ps | -- |

**Supplementary table 1: Sample variation of spin/valley dynamics.** We performed measurements across 6 different samples. All samples were fabricated via mechanical exfoliation followed by dry-transfer without controlling the stacking angle. We observe similar behavior between spin dynamics of electrons and holes across these samples. Our results confirm that stacking order and stacking angle do not significantly affect the spin dynamics in our heterostructures. Sample #3 was covered with drop casted $F_4$TCNQ molecules, which are electron acceptors and induce p-doping in the sample[1]. We observe a slight change in relative spin lifetimes between electrons and holes, and conclude that it will be interesting to investigate spin dynamics in TMD heterostructures under controlled doping.